\documentclass[showpacs, oneside, twocolumn, prl, amsmath, amssymb, nofootinbib, superscriptaddress]
{revtex4-1}
\pdfoutput=1
\usepackage{cases}
\usepackage{amsmath}
\usepackage{amssymb}
\usepackage{amsfonts}
\usepackage{amssymb}
\usepackage{dcolumn}
\usepackage{bm}
\usepackage{bbm}
\usepackage{graphicx}
\usepackage{xcolor}
\usepackage{array}
\usepackage{subfigure}
\usepackage{hyperref}
\usepackage{mathrsfs}

\newcommand{\be}{\begin{equation}}
\newcommand{\ee}{\end{equation}}
\newcommand{\ba}{\begin{eqnarray}}
\newcommand{\ea}{\end{eqnarray}}

\newcommand{\gsim}{\mathrel{\hbox{\rlap{\lower.55ex \hbox {$\sim$}}
			\kern-.3em \raise.4ex \hbox{$>$}}}}
\newcommand{\lsim}{\mathrel{\hbox{\rlap{\lower.55ex \hbox {$\sim$}}
			\kern-.3em \raise.4ex \hbox{$<$}}}}

\hypersetup{colorlinks=true,
	breaklinks=true,
	pdfstartview=Fit,
	linkcolor=blue,
	citecolor=blue,
	urlcolor=blue}

\begin{document}

\title{Sound speed resonance of the stochastic gravitational wave background}

\author{Yi-Fu Cai}
\email{yifucai@ustc.edu.cn}
\affiliation{Department of Astronomy, School of Physical Sciences, University of Science and Technology of China, Hefei, Anhui 230026, China}
\affiliation{CAS Key Laboratory for Researches in Galaxies and Cosmology, University of Science and Technology of China, Hefei, Anhui 230026, China}
\affiliation{School of Astronomy and Space Science, University of Science and Technology of China, Hefei, Anhui 230026, China}

\author{Chunshan Lin}
\email{chunshan.lin@uj.edu.pl}
\affiliation{Faculty of Physics, Astronomy and Applied Computer Science, Jagiellonian University, 30-348 Krakow, Poland}

\author{Bo Wang}
\email{ymwangbo@ustc.edu.cn}
\affiliation{Department of Astronomy, School of Physical Sciences, University of Science and Technology of China, Hefei, Anhui 230026, China}
\affiliation{CAS Key Laboratory for Researches in Galaxies and Cosmology, University of Science and Technology of China, Hefei, Anhui 230026, China}
\affiliation{School of Astronomy and Space Science, University of Science and Technology of China, Hefei, Anhui 230026, China}

\author{Sheng-Feng Yan}
\email{sfyan22@mail.ustc.edu.cn}
\affiliation{Department of Astronomy, School of Physical Sciences, University of Science and Technology of China, Hefei, Anhui 230026, China}
\affiliation{CAS Key Laboratory for Researches in Galaxies and Cosmology, University of Science and Technology of China, Hefei, Anhui 230026, China}
\affiliation{School of Astronomy and Space Science, University of Science and Technology of China, Hefei, Anhui 230026, China}

\begin{abstract}
We propose a novel mechanism to test time variation of the propagation speed of gravitational waves (GWs) in light of GWs astronomy. As the stochastic GWs experience the whole history of cosmic expansion, they encode potential observational evidence of such variation. We report that, one feature of a varying GWs speed is that the energy spectrum of GWs will present resonantly-enhanced peaks if the GWs speed oscillates in time at high-energy scales. Such oscillatory behaviour arises in a wide class of modified gravity theories. The amplitude of these peaks can be at reach by current and forthcoming GWs instruments, hence making the underlying theories falsifiable. This mechanism reveals that probing the variation of GWs speed can be a promising way to search for new physics beyond general relativity.
\end{abstract}

\maketitle

%\noindent\textbf{Introduction.}
{\it Introduction.} --
The current understanding of nature is based on the theory of general relativity (GR) and the standard model (SM) of particle physics. While these theories remain extremely successful, it is crucial to question how one can probe new physics beyond them. One approach is to search for observable effects of possible deviations from the constants on which the theory depends. In particular, the sound speed of gravitational waves $c_g$ is a fundamental constant in the context of GR. It characterizes how rapidly a change in the distribution of energy and momentum of matter or of the gravitational field itself results in subsequent alteration of the gravitational field over a distance in spacetime. In GR, GWs propagate at the same speed as light $c$. This was confirmed with high precision $-3 \cdot 10^{-15} \leq c_{g} / c-1 \leq 7 \cdot 10^{-16}$ at low redshift by the multi-messenger observation GW170817/GRB170817A \cite{Monitor:2017mdv}. However, the observational limits are considerably weaker at large scales. Hence, in principle $c_g$ could deviate from $c$ at large scales which would constitute an indication of new physics beyond GR \cite{Ezquiaga:2018btd}. Recently, the North American Nanohertz Observatory for Gravitational Waves (NANOGrav) experiment has reported results of the search for stochastic background of GWs based on its 12.5-year data set \cite{Arzoumanian:2020vkk}, which might indicate that new physics beyond GR can be probed with GW experiments in the near future.

In this Letter, we investigate what happens if the sound speed of GWs $c_g$ oscillates at primordial era. In physical systems, parametric resonance can occur if some parameters or degrees of freedom oscillate. This mechanism is widely studied and applied in condensed matter physics \cite{hernandez2007parametric, brau2011multiple}, and in cosmological setups, namely, cosmic preheating \cite{Traschen:1990sw, Dolgov:1989us, Kofman:1997yn, Allahverdi:2010xz}, sound speed resonance (SSR) \cite{Cai:2018tuh, Cai:2019jah, Chen:2019zza, Chen:2020uhe}, and so on. As primordial fluctuations can be resonantly amplified via SSR, the corresponding density perturbations can lead to the formation of primordial black holes (PBHs) in post-inflationary phases \cite{Cai:2018tuh}, and a stochastic background of GWs can be significantly induced through nonlinear process \cite{Cai:2019jah, Zhou:2020kkf}. Thus, a crucial question follows: can GWs production become more efficient when the resonance takes place in the gravitational sector itself? For instance, a broad resonance for primordial GWs could occur if graviton obtains an oscillating mass \cite{Lin:2015nda, Kuroyanagi:2017kfx}. In this Letter, we report that GWs can be considerably enhanced in the very early universe via narrow resonance, and that the underlying new physics can be probed by GW astronomy.

%\noindent\textbf{Theoretical background.}
{\it Theoretical background.} --
According to terrestrial experiments, new physics beyond the SM of particle physics is expected to kick in at energy scales above $\sim 13$TeV \cite{1811596}. It happens that the very early universe naturally provides us an ideal environment at high energy scales to discover possible hints of new physics beyond both SM and GR such as from, for instance, quantum gravity theories \cite{Rovelli:2000aw}. Concerning  gravitation, modified gravity (MG) is one of the explicit and classical way for probing new physics beyond GR, which has been extensively studied and has been found to yield fruitful cosmological phenomenons including dark energy, GWs and the physics in the early universe \cite{Copeland:2006wr, Ezquiaga:2018btd}, such as torsion gravity \cite{Cai:2015emx, Cai:2018rzd, Li:2018ixg, Yan:2019gbw}, scalar-tensor theories \cite{Langlois:2015cwa, Heisenberg:2018vsk, Ilyas:2020qja, horndeski, Deffayet:2011gz, Kobayashi:2011nu, Gleyzes:2014dya, Lin:2014jga, Langlois:2015cwa, Gao:2014soa}, massive gravity \cite{Fierz:1939ix, deRham:2010kj}, etc.

The oscillatory sound speed of GWs can arise in a wide class of MG theories. For instance, in Horndeski theories, the sound speed of GWs takes $c_g^2=\mathcal{F}_T/\mathcal{G}_T$ \cite{Kobayashi:2019hrl}, where $\mathcal{F}_T$ and $\mathcal{G}_T$ are functions of scalar field and its temporal derivative that are determined by the underlying theory. If we assume that this scalar field dominates over the universe during reheating, its coherent oscillation can then give rise to an oscillatory behaviour in $c_g^2$. For the 4-dimensional Einstein-Gauss-Bonnet (EGB) gravity \cite{Glavan:2019inb}, $c_g^2$ receives a correction proportional to $\dot{H}/M_p^2$ with $H$ being the Hubble parameter and the dot representing time derivative. As such, it can also be oscillatory around the reheating epoch. Similar oscillatory corrections can be realized in many scalar-tensor theories beyond Horndeski \cite{Gleyzes:2014dya, Lin:2014jga, Langlois:2015cwa, Kobayashi:2019hrl}. Without loss of generality, we in this Letter perform a model-independent analysis by parameterizing the sound speed of GWs as,
\be \label{Eq:CsParam}
 c_g^2 = 1 - \frac{\alpha}{(1+\tau/\tau_0)^2} \cos^2(k_{*}\tau) ~.
\ee
Here $\tau$ is conformal time, and $\tau_0$ and $k_*$ are the characteristic time scale and wave number, respectively. We demand $\alpha>0$ to avoid superluminal propagation\footnote{It was argued in \cite{deRham:2019ctd, deRham:2020zyh, Alberte:2020jsk} that the subluminality in the GWs sector may also lead to the causality violation. Nevertheless, we mention that our results are insensitive to the sign of $\alpha$, namely flipping the sign of $\alpha$ only shifts a phase inside the cosine function of Eq.~\eqref{Eq:Mathieu} and thus does not change our main conclusions.}. The specific time dependence form in Eq.~\eqref{Eq:CsParam} is well motivated, as we would generally expect the scalar field to oscillate about its minimum during reheating. This yields a cosine factor, while the amplitude of the oscillation decreases due to the Hubble friction as the universe expands, which leads to a time-dependent factor in the denominator of Eq.~\eqref{Eq:CsParam}.
{For simplicity, we assume that the corrections to the background are negligible and thus background evolution is the one as in standard $\Lambda$CDM cosmology. From the effective field theory (EFT) perspective \cite{Cheung:2007st}, this type of models could be achieved by assuming an oscillatory feature in the time evolution of some EFT operators.}
In passing, we comment that $c_g^2$ reduces to unity in the late-time limit $\tau\to\infty$, and thus satisfies the constraint from the multi-messenger observation GW170817/GRB170817A.

The oscillatory correction term in Eq.~\eqref{Eq:CsParam} leads to a narrow parametric resonance at the sub-Hubble scale, given proper values to the model parameters. It is of observable interest if this scale corresponds to the frequencies of current or near future GW detectors. We are thus more interested in the sub-Hubble modes, where $k \gg {a^{\prime}}/{a}$ with $a$ being the scale factor of the universe and the prime being the derivative with respect to conformal time. At this scale, we can safely neglect the Hubble friction term, as well as the possible mass term, which stems from the canonical normalization of the GWs quadratic action. Hence, by introducing $x=k_*\tau$, the equation of motion for GWs can be approximately expressed in form of the Mathieu equation,
\be \label{Eq:Mathieu}
 \frac{\partial^{2} h_{k}}{\partial x^{2}}+[A-2 q \cos (2 x)] h_{k} = 0 ~,
\ee
where $A=\ {k^2}/{k_*^2}-2q$ and $q = \alpha k^2 / [4k_*^2 (1+\frac{x}{x_0})^{2}]$. The primary characteristic of the solutions is an exponential instability $h_k\propto\text{exp}(\mu_k^{(n)}x)$ within certain resonance bands, where $\mu_k^{(n)}$ is called the Floquet Exponent. The first resonance band is the most efficient, which appears between $A\in (1-q, 1+q)$ with $\mu_k^{(1)}\simeq q/2$ for $q\ll 1$. Due to the cosmic expansion, a mode enters the resonance band, undergoes the exponential growth for a short period, and then automatically exits this band. Thus, the amplification of GWs can become controllable.

{\it Parametric resonance.} --
In present study, we take inflation can set initial conditions for primordial GWs. Quantum fluctuations of spacetime appear at sub-Hubble scales, so that they behave as plane waves with $h_{k}(\tau) \cdot {a M_\text{Pl}}/{2 \sqrt{2}} = e^{-i c_s k \tau} / \sqrt{2 c_s k}$, with $c_s$ being the sound speed of GWs during inflation. Hence, it is not necessarily the same as the parameterization given in Eq.~\eqref{Eq:CsParam}. Note that, for a broad class of MG theories, $c_s$ generally deviates from unity due to the presence of possible high-order derivative terms. Moreover, the quasi-exponential expansion stretches the modes out of the Hubble horizon, which eventually yields the power spectrum: $\Delta_{t}^{2}(k, \tau)=\frac{k^{3}}{2 \pi^{2}}\left|h_{k}(\tau)\right|^{2} \simeq {2}\left({H} / {c_s M_\text{Pl}}\right)^{2} \left(c_sk\right)^{n_t} / {\pi^{2}}$, where $n_t$ is the spectral index.

\begin{figure*}[ht]
\centering
\subfigure{\includegraphics[width=2.8in]{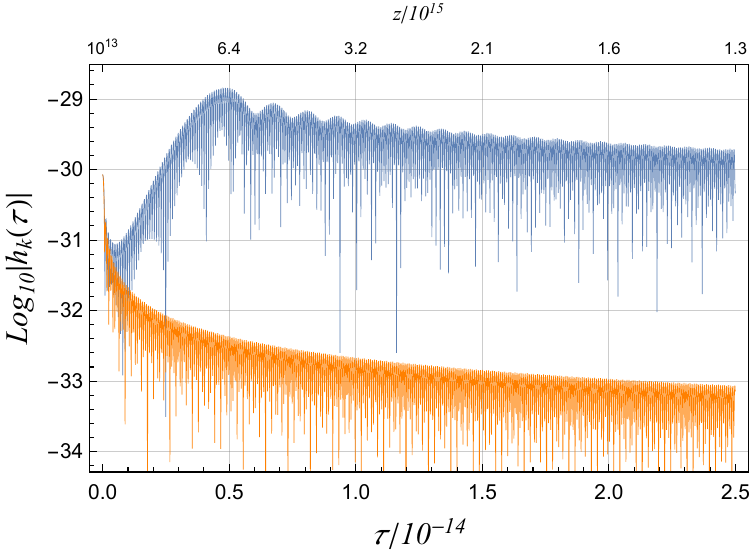}}
\subfigure{\includegraphics[width=2.8in]{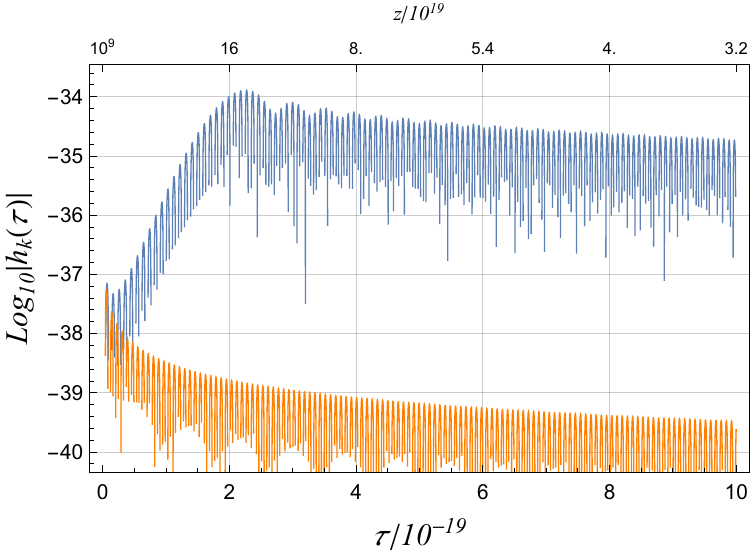}}
\subfigure{\includegraphics[width=2.8in]{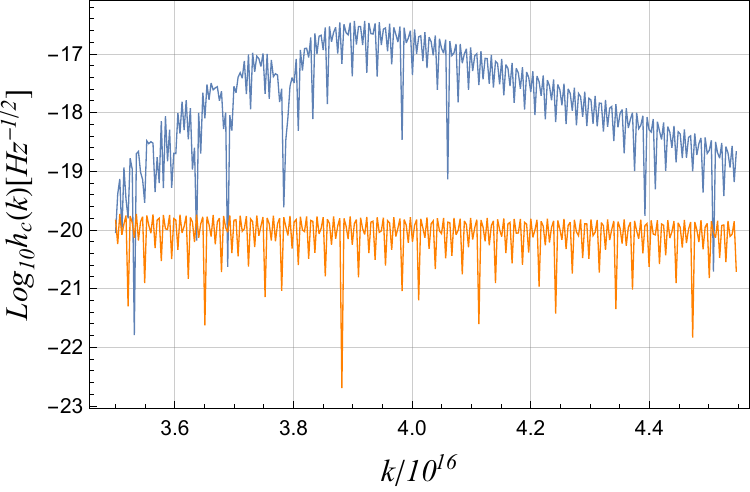}}
\subfigure{\includegraphics[width=2.8in]{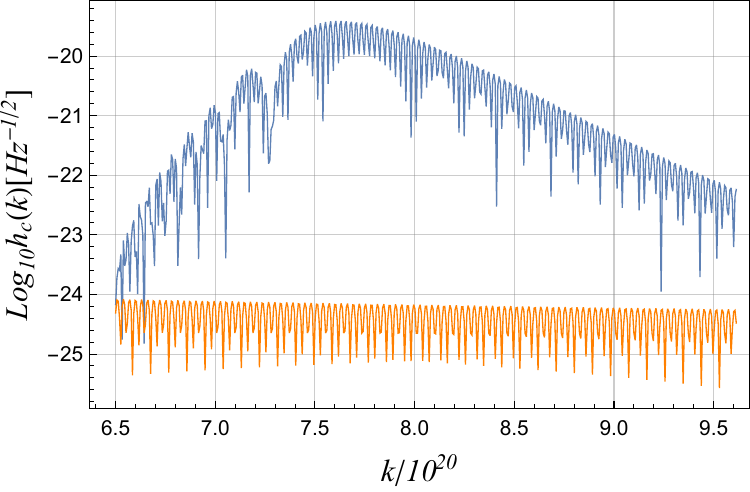}}
\caption{Upper panel: GWs $h_k(\tau)$ with respect to conformal time $\tau$ (or to redshift $z$ as shown on the top). Lower panel: Characteristic spectra $h_{c}(k)$ with respect to $k$. For all panels, the pale blue and orange curves denote the numerical results with ($c_g^2$ in Eq.~\eqref{Eq:CsParam}) and without ($c_g^2=1$) resonance, respectively. In left panel, we set $k_{*}=3.5 \times 10^{16}$ and $k=3.9 \times 10^{16}$ for which space-based detectors are sensitive to, and deviation parameter $\alpha=0.7$, and characteristic time $\tau_{0}=7.5 \times 10^{-15}$.  In right panel, we set $k_{*}= 6.5 \times 10^{20}$, $k = 6.8 \times 10^{20}$ for which terrestrial detectors are sensitive to, and $\alpha = 0.95$, $\tau_{0} = 3.7 \times 10^{-19}$.}
\label{fig:ModesEvolution}
\end{figure*}

The parametric resonance, which is triggered by the coherent oscillation of a scalar in certain MG theories, may occur during or after reheating epoch. As we have explained, the modes could enter and exit the resonance bands due to the cosmic expansion. Thus, Eq.~\eqref{Eq:Mathieu} is oversimplified for the physics that we are interested. A full picture requires us to additionally consider the redshift term in the equation of motion of GWs,
\be \label{Eq:EOMtensor}
h_{k}^{\prime \prime}(\tau) + 2 \mathcal{H} h_{k}^{\prime}(\tau) + c_g^2 k^{2} h_{k}(\tau) = 0 ~,
\ee
where $\mathcal{H}$ is the conformal Hubble parameter $\mathcal{H}\equiv a^{\prime}(\tau)/a(\tau)$.
{Note that, besides the oscillatory sound speed of GWs, we will neglect other possible MG effects and work within the inflationary $\Lambda$CDM cosmology.}
For an accelerating background, $a(\tau) = l_H |\tau-\tau_a|^{-\gamma}$, with the normalization $|\tau_H-\tau_a|=1$ and $l_H\equiv \gamma/H_0$ \cite{Zhang:2006mja, Zhang:2010er}, where $\tau_H=1.101$ is the present conformal time, and $\gamma=2.265$, $\tau_a=2.101$, $H_0=67.4\, {\rm km}\,{\rm s}^{-1}{\rm Mpc}^{-1}$ according to CMB observation \cite{Aghanim:2018eyx}. The conformal wave number $k$ is related to the physical frequency at $\tau$ via $f(\tau) =  k/ \big( 2\pi a(\tau) \big)$. This generalized Mathieu equation can be solved numerically
{throughout the whole evolution of the universe. Here the cosmological parameter values are inferred by Planck \cite{Aghanim:2018eyx}.}
The results are reported in Fig.~\ref{fig:ModesEvolution}. At the very beginning of radiation domination, the initial amplitudes of sub-Hubble modes were originally set by those produced during inflation and thus our analysis is consistent with the initial condition discussed previously. In the upper panel of Fig.~\ref{fig:ModesEvolution}, we present the evolution of stochastic GWs under parametric resonance by choosing two sets of parameter values. Accordingly, there are two characteristic frequency bands with one being sensitive to space-based detectors (left panel) and the other terrestrial detectors (right panel).

We show that, resonance continuously enhances stochastic GWs background until $\tau \sim \tau_0$. For space-based detectors, we are concerned with the mode around $k=3.9 \times 10^{16}$, which corresponds to the frequency $f=0.0066 \mathrm{Hz}$ at present just within the sensitive range of Laser Interferometer Space Antenna (LISA). The mode around $k=6.8 \times 10^{20}$ corresponding to the frequency $f=115 \mathrm{Hz}$, are instead sensitive to terrestrial experiments, such as the Laser Interferometer Gravitational-Wave Observatory (LIGO), Einstein Telescope (ET), and Cosmic Explorer (CE). In general, the amplitude of stochastic GWs background decays dramatically with time because Hubble expansion. However, for enhanced modes, resonance sustains about $\Delta\tau \sim 10^{-15}$ around $\tau_{0}=7.5 \times 10^{-15}$ for space-based surveys and $\Delta\tau \sim 10^{-19}$ around $\tau_{0}=3.7 \times 10^{-19}$ for terrestrial experiments. Each of them contributes to $10^4$ and $10^5$ times enhancement due to exponential factor $\text{exp}(\alpha k_{*} \Delta \tau / 16)$, respectively.

Moreover, one often confronts the characteristic spectrum $h_c(f)$ with observations as $h_c(f) \equiv {h(k,\tau_H)} / {(2\sqrt{f})}$, where the spectrum is also written as $h(f,\tau_H ) \equiv \sqrt {\Delta^2_t(k,\tau_H) }$ \cite{Maggiore:1999vm, Zhang:2018dvc}. Thus, we illustrate the characteristic spectra of the above two modes evolving to the end of radiation domination by displaying their first resonance bands in the lower panel of Fig.~\ref{fig:ModesEvolution}. Due to the property of narrow resonance, these bands span a very narrow interval $\Delta k \sim q k_*$ on the spectra. Also, the amplification magnitudes for both cases are consistent with their evolution along $\tau$.

\begin{figure*}[ht]
\centering
\subfigure{\includegraphics[width=2.8in]{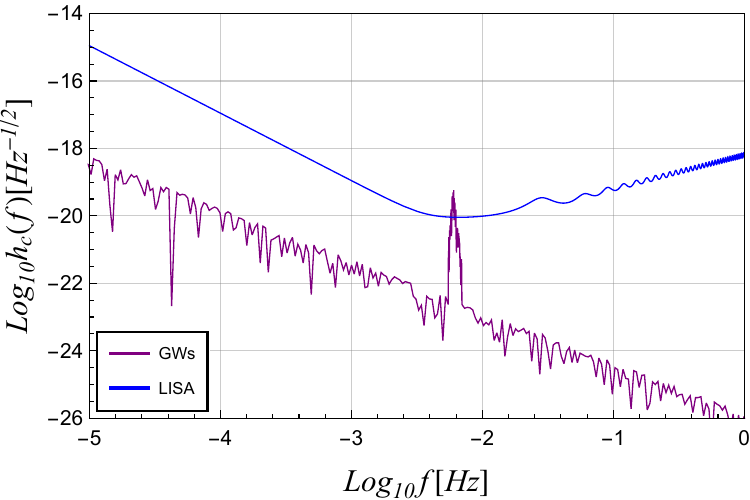}}
\subfigure{\includegraphics[width=2.8in]{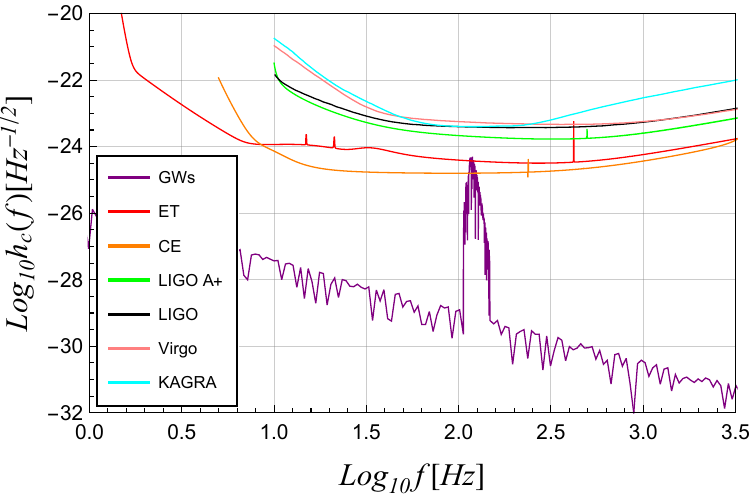}}
\subfigure{\includegraphics[width=2.2in]{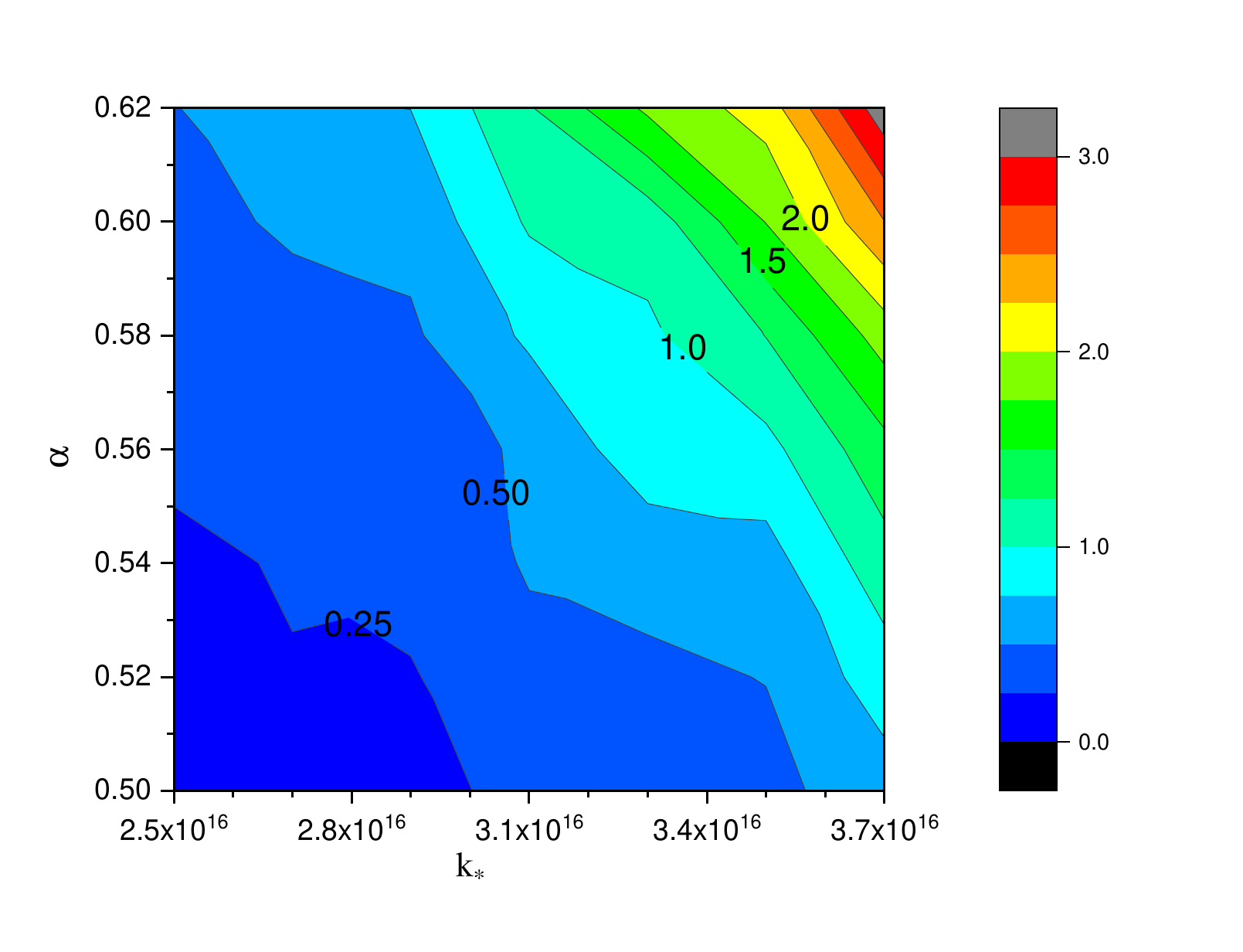}}
\subfigure{\includegraphics[width=2.2in]{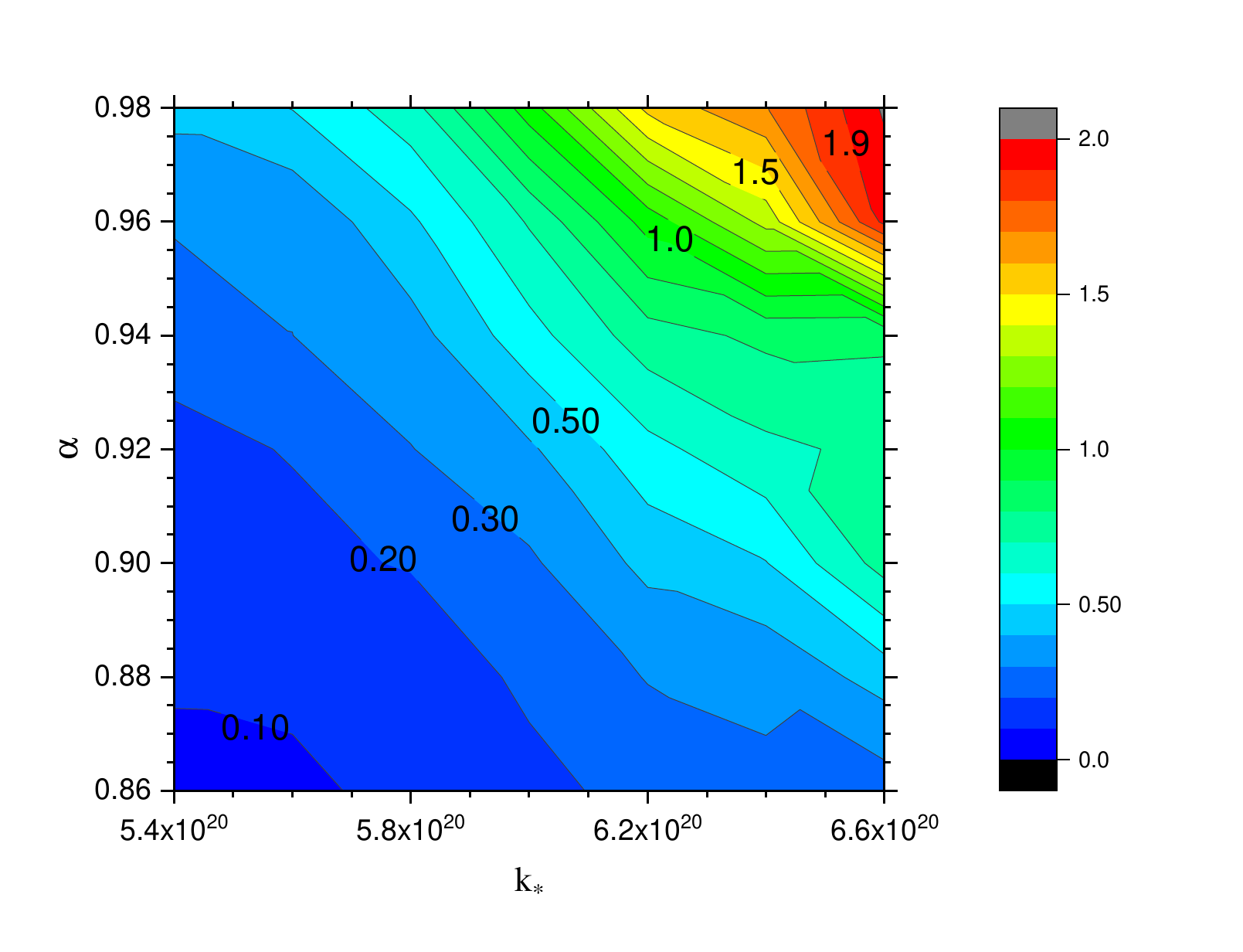}}
\subfigure{\includegraphics[width=2.2in]{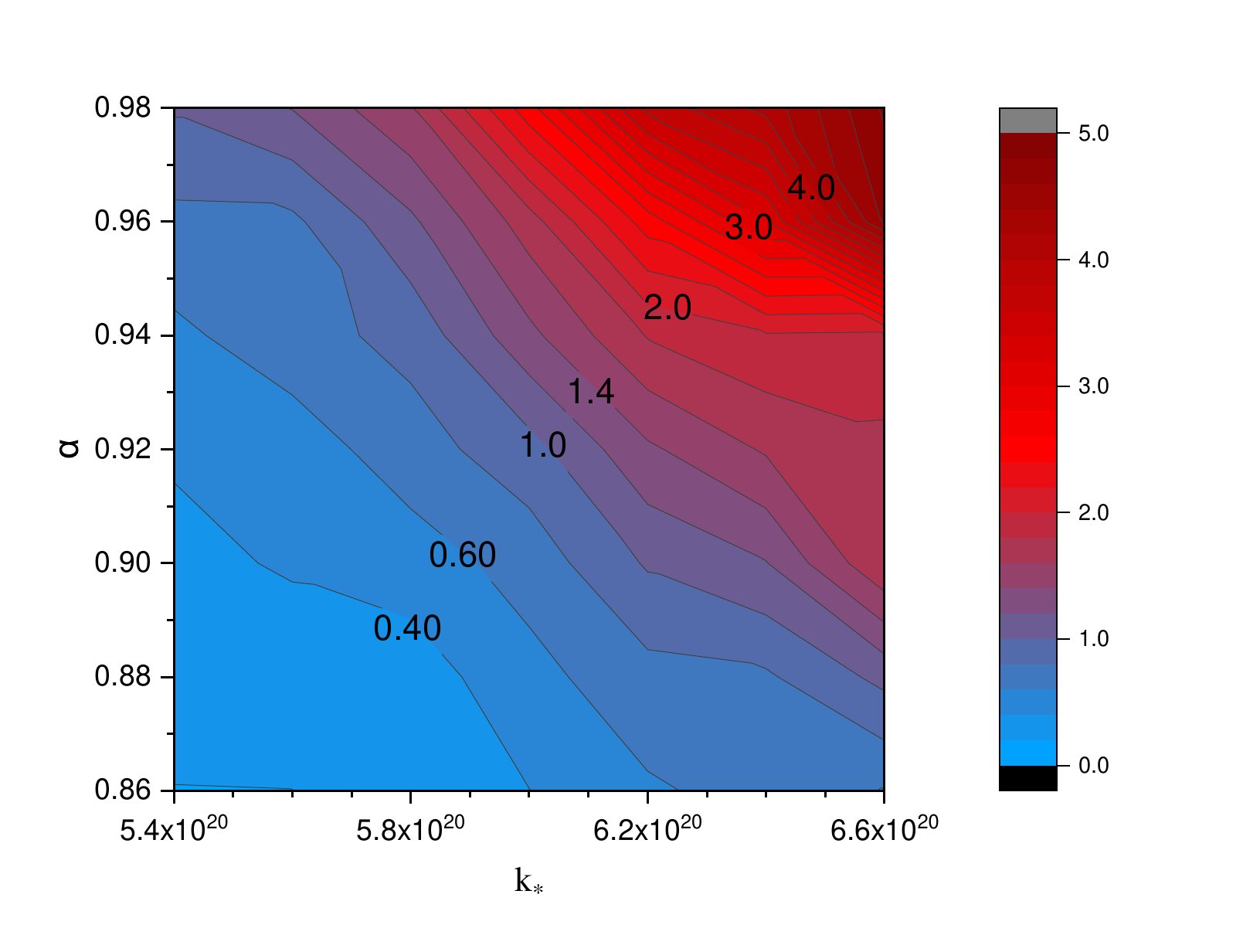}}
\caption{Upper panel: GWs signals (purple curves) versus experimental noise curves as a function of $k$. The values of model parameters are the same as in Fig.~\ref{fig:ModesEvolution}. Left upper panel shows the noise curve of LISA (blue) \cite{Larson:1999we}. Right upper panel shows the noise curves of ET (red), CE (orange) \cite{Dwyer:2014fpa, Evans:2016mbw}, LIGO A+ (green) \cite{LIGOaPlusNoise}, LIGO (black), Virgo (pink) and Kamioka Gravitational-Wave Detector (KAGRA) (cyan) \cite{Abbottetal2016LIGOdesignnoise, Aasi:2013wya}, respectively.
Lower panel: SNR distributions with respect to $\alpha$ and $k_*$ for LISA (left), ET (middle) and CE (right), respectively.}
\label{fig:Observations}
\end{figure*}

{\it Observational constraints.} --
As shown in Fig.~\ref{fig:Observations}, it is straightforward to evaluate GWs to present and make comparison to the noise curves of experiments. From the left upper panel, by adopting the same parameter set in modes evolution, the peak on characteristic spectrum with almost $10^4$ times amplification is detectable to LISA. We find that the most intense resonance takes place at redshift $z \simeq 6 \times 10^{15}$, which corresponds to $\mathcal{H}=1.7 \times 10^{14} \ll k \sim 10^{16}$. This confirms that the approximation we made in Eq.~\eqref{Eq:Mathieu} is valid for semi-analytical calculations. Interestingly, the energy scale of the universe around that redshift is $\sim \text{TeV}$. For terrestrial experiments, looking at the right upper panel of Fig.~\ref{fig:Observations}, the spectrum is enhanced about 5 orders of magnitude with a peak around characteristic frequency. In this case, the most intense resonance takes place at $z \simeq 1.2 \times 10^{20}$, which corresponds to $\mathcal{H}=3 \times 10^{18} \ll k \sim 6 \times 10^{20}$. This is consistent with the approximation of Eq.~\eqref{Eq:Mathieu}. The corresponding energy scale is $\sim 10^4$ TeV. These features indicate that a broad class of scalar-tensor theories could leave us observable footprint of new physics on stochastic GWs background spectrum given that the early universe energy scales are greater than $\sim$ TeV.

We observe that the peak of such spectrum is within an extremely narrow band of the frequency range, and therefore it can easily satisfy most astrophysical bounds. However, the signal-to-noise ratio (SNR) provides the lower bound for model parameters through
\be \label{eq:SNR}
 \text{SNR} \equiv \frac{h_{c} \left(f\right)}{h_{\mathrm{detector}} \left(f\right)} ~,
\ee
where $h_{\mathrm{detector}}(f)$ is the noise curve for GWs detectors. To illustrate the detectable parameter space, the SNR distributions with respect to $\alpha$ and $k_*$ are shown in the lower panel of Fig.~\ref{fig:Observations} for LISA, ET and CE, respectively. We take $\mathrm{SNR}>1$ as the threshold for detectable stochastic GWs background signals. For LISA, we set $\mathrm{SNR} = {h_{c}(f_{0})} / {h_{\mathrm{LISA}}(f_{0})}$, and $f=f_{0}$ maximizes the value of ${h_{c}(f)} / {h_{\text {LISA }}(f)}$. Meanwhile, we define $\mathrm{SNR}=h_{c}(f_{0}) / h_{\mathrm{ET}}(f_{0})$ for ET and $\mathrm{SNR}=h_{c}(f_{0}) / h_{\mathrm{CE}}(f_{0})$ for CE, where $f=f_{0}$ maximizes the value of ${h_{c}(f)} / {h_{\text {LIGO }}(f)}$. According to the contours in Fig.~\ref{fig:Observations}, we find that there is a detection window for LISA, ET, and  CE on the proposed mechanism, through probes of SSR signals of the stochastic GWs background.

{\it Nonlinear effects on density perturbations.} --
Previously, we have shown that MG theories at the electroweak scale may give rise to a sharp peak on the stochastic GWs spectrum via SSR. Thanks to the cosmic expansion, all modes that can enter the resonance band have to exit eventually. Accordingly, the enhancement factor as well as the frequencies of the resonance peaks are controllable. Note that the LISA sensitivity range is often relevant for PBHs formation. Hence, it is intriguing to further ask whether the sharp peak on the stochastic GWs spectrum at this scale may lead to an extra contribution to density perturbations so that it might be related to the PBHs formation, provided that the resonance amplification is large enough for the nonlinear effects to become crucial. In our case, it is possible for the peak of tensor modes to generate the overdense region at the nonlinear level without spoiling the cosmological perturbation theory, provided that (1) the peak is high enough, yet still less than unity, and (2) the scalar-tensor-tensor coupling is sizable. At the sub-Hubble scale, GWs can source scalar modes at the nonlinear level, $\zeta'' +2\mathcal{H}\zeta' -\partial^2\zeta= \lambda_{stt} \partial_i h_{jk} \partial_i h_{jk} +...$, where $\lambda_{stt}$ is the scalar-tensor-tensor coupling coefficient. In Fourier space, the source term acquires the convolution over the whole momentum range. However, the most significant contribution comes from the resonance peak. Hence, the scalar mode can be estimated as $\zeta_{k_*} \sim \lambda_{stt}|h_{k_*}^2|$, where $|h_{k_*}^2|$ is the amplitude of the resonance peak of the GWs spectrum. The amplitude is bounded by BBN constraint $\rho_{GW}\simeq \frac{1}{32\pi G}\frac{k_*^2}{a_{BBN}^2}|h_{k_*}^2|<0.05 \rho_\gamma$, where $\rho_\gamma$ is the radiation energy density. The Mathieu equation requires $k_*^2> \mathcal{H}^2$ to have the narrow parametric resonance, and thus $|h_{k_*}^2|<10^{-1}$ is bound from above. Provided that the BBN bound is satisfied, a peak with amplitude around $10^{-2}$ on scalar modes at the same scale could be achieved if $\lambda_{stt}>1$, and consequently might lead to PBHs formation. To our knowledge, this is the first time such scenario is discussed. More investigation along these lines will be addressed in follow-up studies.

{\it Conclusions.} --
In this Letter we reported a novel mechanism that can produce detectable GW signals from a resonantly-enhanced stochastic GWs spectrum due to an oscillatory sound speed. This mechanism bridges GWs astronomy with new physics beyond GR that may appear above electroweak scales. Such new physics may arise from a broad class of scalar-tensor theories that could yield an oscillating $c_g^2$. The enhancement of GWs spectra is manifestly different from the SSR for scalar perturbations, because the oscillatory correction for $c_g^2$ is time-varying due to the nature of cosmic expansion. It turns out that, $c_g^2$ approaches unity in the late-time universe where GR can be recovered. Our result reveals that the energy spectrum of stochastic GWs presents a peak feature with an enhanced amplitude that can be at reach for current and forthcoming terrestrial (LIGO, Virgo, LIGO A+, CE, ET and KAGRA) and space-based (LISA) GWs surveys. Moreover, we selectively performed the SNR analyses of LISA, CE and ET to illustrate that our mechanism is viable in probing new physics beyond GR. 
{Additionally, our mechanism can be straightforwardly extended by including anisotropic stress, high-order damping terms and other possible MG effects, making them testable in the GW astronomy.}

We note that the TeV energy scale, on which gravity is possibly modified, might be accessible for LHC and next-generation colliders. We are prompted to ask whether the modifications also lead to some new phenomenons for collider physics. If one introduces the coupling $\phi^2 {\cal H}^\dagger {\cal H}$ (unless the shift symmetry of $\phi$ prohibits it) where $\phi$ arises from the MG sector and ${\cal H}$ is the Higgs boson in the SM of particle physics, the particle collisions could lead to the excitation of the scalar boson $\phi$, which would subsequently decay to graviton as a tertiary product. However, the frequency of GWs produced via this process shall be around the TeV scale, which remain challenging to be probed by experiments.

We end by commenting that we also put forward a novel mechanism of enhancing density perturbations by virtue of the nonlinear effects from scalar-tensor-tensor couplings. While this is a byproduct of the present study, it deserves a detailed analysis in a follow-up project, in particular, regarding the possibility of PBHs formation. Additionally, we point out that the nonlinear coupling among three tensor modes could also bring extra contribution to the stochastic GWs background by yielding a relatively flat spectrum within low frequency bands which could be examined by pulsar timing array surveys, 
{such as the recently released NANOGrav data.}

%\noindent\textbf{Acknowledgments.}
{\it Acknowledgments.--}
We are grateful to R. Brandenberger, D. Glavan, M. Sasaki, S. Pi, Y. Wang, M. Yamaguchi and P. Zhang for valuable discussions. This work is supported in part by the NSFC (Nos. 11961131007, 11653002, 11722327, 1201101448, 11421303, 12003029), by the CAST-YESS (2016QNRC001), by the National Youth Talents Program of China, and by the Fundamental Research Funds for Central Universities.
CL is supported by the grant No. UMO-2018/30/Q/ST9/00795 from the National Science Centre (Poland).
BW is also supported by the China Postdoctoral Science Foundation grant No. 2019M662168.
All numerics were operated on the computer clusters {\it LINDA} \& {\it JUDY} in the particle cosmology group at USTC.

\end{document}